\documentclass[useAMS,usenatbib]{mn2e}
\usepackage{graphicx}

\title[The properties of friends-of-friends groups]
{Groups in the Millennium
  Simulation and in SDSS DR7}

\author[Nurmi P. et al.]{P. Nurmi$^{1}$\thanks{pasnurmi@utu.fi}, P.
Hein\"am\"aki$^{1}$, T. Sepp$^{2,5}$, E. Tago$^{2}$, E. Saar$^{2,4}$, M. Gramann$^{2}$, 
\and M. Einasto$^{2}$, E. Tempel$^{2,3}$ and  J. Einasto$^{2,4,6}$\\
$^{1}$Tuorla Observatory, Department of Physics and Astronomy, 
University of Turku, V\"ais\"al\"antie 20, FI-21500 Piikki\"o,
Finland\\
$^{2}$Tartu Observatory, T\~oravere, Tartumaa, 61602 Estonia\\
$^{3}$National Institute of Chemical Physics and Biophysics, Tallinn 10143, Estonia\\
$^{4}$Estonian Academy of Sciences,  EE-10130 Tallinn, Estonia\\
$^{5}$Institute of Physics, University of Tartu, Estonia\\
$^{6}$ICRANet, Piazza della Repubblica 10, 65122 Pescara, Italy
}

\begin{document}
\date{Accepted -. Received -; in original form -}
\pagerange{\pageref{firstpage}--\pageref{lastpage}} \pubyear{2013}
\maketitle

\label{firstpage}

\begin{abstract}
The Millennium N-body simulation and SDSS DR7 galaxy and galaxy group
catalogues are compared to study the properties of galaxy groups and
 the distribution of galaxies in groups.
We construct mock galaxy group catalogues for a Millennium
semi-analytical galaxy catalogue by using the same friends-of-friends
method, which was used by Tago et al. (2010) to analyse the SDSS data.
 We analyse in detail the group luminosities, group richnesses, virial radii, sizes
of groups and their rms velocities
for four volume-limited samples from observations and simulations.
Our results show that the spatial densities of groups agree within one order of magnitude in
all samples with a rather good agreement between the mock catalogues
and observations.
All group property distributions have similar shapes and
amplitudes for richer groups. For galaxy pairs and small groups the group
properties for observations and simulations are clearly
different.
In addition, the spatial distribution of galaxies in small groups is different:
at the outskirts of the groups the galaxy number distributions do not agree,
although the agreement is relatively good in the inner regions.
Differences in the distributions are mainly due to the observational
limitations in the SDSS sample and to the problems in the semi-analytical methods that
produce too compact and luminous groups.
\end{abstract}

\section{Introduction}

In the hierarchical picture of galaxy formation, galaxies form by 
baryon cooling within dark matter (DM) haloes; after that they cluster gravitationally to
form galaxy groups and clusters of galaxies. During this process several 
hydrodynamical (e.g., ram pressure stripping, viscous stripping, strangulation)  
and gravitational (tidal interaction between galaxies,
galaxy merging) processes together lead to  
virialized groups and clusters. At the same time these processes modify
galaxy morphologies in dense environments and establish the well known 
morphology-density relation (Einasto et al. 1974, Dressler \citeyear{dress}, Potsman \& Geller \citeyear{pots}). 
In the hierarchical picture,   
groups and clusters of galaxies are generally assumed to be systems 
embedded in an extended dark matter halo whereas  
satellite galaxies themselves reside in dark
matter subhaloes. Simulations strongly support this picture, 
although still some problems related to the smallest
DM subhaloes remain (e.g., Guo et al. \citeyear{guo}, Klypin et al. \citeyear{klypin11}). 

Recent numerical and analytical studies
of the cluster scale DM haloes show that cluster abundances (Press \& Schechter \citeyear{press},
Jenkins et al. \citeyear{jen:jen} and Sheth \& Tormen
\citeyear{sheth}) and two-point correlation functions agree well
(Springel et al. \citeyear{springel}, Conroy, Wechsler and Kravtsov \citeyear{conroy}) 
with observations. 
When the galaxy
formation physics, incorporated in smoothed-particle hydrodynamics (SPH) simulations, has been taken into
account,
good agreement with observed galaxy clustering is obtained 
(e.g., Weinberg et al. \citeyear{weinberg} and Maller et al. \citeyear{maller}).
Summarizing, the agreement
between the observational data and theoretical studies of
galaxy clusters is relatively good, but is the agreement as
good for the galaxy group scale?

From observations we know that most galaxies are situated in galaxy 
groups (e.g. Geller \& Huchra 1983, Mulchaey \citeyear{mul2000}, Eke et
al. \citeyear{eke}, Karachentsev \citeyear{kara}), meaning that group
environment has an important role in structure formation and 
galaxy evolution.  
Different group catalogues can be compiled, even from the same data
set, by using different algorithms. The most widely used method is the
friends-of-friends (FOF) algorithm used already by Geller \& Huchra (1983).
Different implementations of this algorithm have been used, e.g., for the
following group catalogues: Tucker et al. (2000), Allam \&
Tucker (2000), Blanton et al. (2005), Berlind et al. (2006), 
Yang et al. (2007), Tago et al. (2008), Tempel et al. (2012), Mu\~{n}oz-Cuartas \& M\"{u}ller (2012) and Wen et al. (2012).

From observations we know that galaxy clusters and galaxy groups are not distinct classes of objects.
This can be seen in a continuous richness distribution from galaxy pairs to 
rich groups and clusters in observations (Berlind et al. \citeyear{berlind}).
Indeed, in many ways 
groups can be viewed as scaled-down versions 
of clusters; e.g., many X-ray scaling relations extend from clusters to 
groups although the scatter in 
the correlations increases towards smaller systems
(Mulchaey \citeyear{mul2000}, Sun et al. \citeyear{sun},
Eckmiller, Hudson and Reiprich \citeyear{eck}).
On the other hand, many studies have found systematic
differences between the physical properties of groups and galaxy clusters 
(Eckmiller et al. \citeyear{eck} and references therein).

The purpose of this paper is to statistically compare properties of groups
in observations and simulations and to find the most important
differences between them. This analysis provides the stepping stone to
more detailed studies of individual group properties that can give more strict
constraints for the semi-analytical methods (SAM) that incorporate presently all complicated physics related to galaxy
formation. Several SAMs have been
applied to the Millennium Run to construct galaxies by using the dark matter merger
trees in the simulation. Our study is based on the galaxies produced by the
semi-analytical procedures by Bertone, De Lucia and Thomas (2007) and
Font et al. (2008).
Using the mock group catalogue and SDSS groups we study how closely
the FOF-groups in simulations resemble the observed galaxy groups. This will tell us how well the
distribution of galaxies defined on the basis of SAM
represents the real distribution of galaxies in groups. 
Similar comparison studies between galaxy group catalogues and
  SAM results have been done, for example, by Weinmann et al. (2006b), Kimm et
  al. (2009) and Liu et al. (2010).

In Section 2 we present both the observational and mock data sets used in this study and 
outline the procedure used to construct
the group catalogue for the Millennium simulation. We compare the galaxy luminosity functions in
observations and simulations, and describe the general statistical
properties of the galaxies. 
In the next section (Section 3) we analyse in detail the statistical properties of
galaxy groups in the SDSS galaxy group catalogue and in the mock catalogue.
We study the group luminosity functions, group richnesses, rms velocities, virial radii and 
maximum sizes and cross-correlate them between 
the mock catalogues and observations. The radial distribution of
galaxies in the groups is studied before the conclusions and discussion
(Section 4).

\section{DATA}
\subsection{Observations}

As a basis for our analysis we have used groups of galaxies
compiled by \cite{tago2010} for
the Sloan Digital Sky Survey seventh data release (SDSS
DR7)\footnote{http://www.sdss.org/dr7/}, (York et al. 2000, Adelman-McCarthy et al.
2008, Abazazian et al. 2009). This group catalogue was derived from the SDSS DR7
main galaxy sample which contains
697920 galaxies. After the extraction process the total
number of galaxies used was reduced to 583362, covering the
redshift range $0.009 < z < 0.2$ and about 25 percent of the full sky.
A small number of groups ($\sim$2\%) are located close to the
  edges of the SDSS field. Since the number of such groups is so
  small, the possible errors that are caused by this should be also
  very small and they do not influence the results.

To study the properties of galaxy groups we have selected different volume-limited samples
from the DR7 data. By using volume selected samples the comparison with the simulations 
is free of many selection effects and the calculated distributions can be directly compared.
The observed magnitudes of individual galaxies in the SDSS Petrosian
$r$-magnitude band $m_r$ are between 12.5 and 17.77.
These limits are used for the absolute magnitude limits $M_r$.
The absolute magnitudes in the group catalogue correspond to the rest frame
at the redshift $z = 0$.
We select four samples S1, S2, S3 and S4 that have faint absolute magnitude limits 
$-18$, $-19$, $-20$ and $-21$.
In this way all galaxies that have
$L_r>L_{\mathrm{lim}}$ are included in the analysis. The $L_{\mathrm{lim}}$ is simply
$L_{\mathrm{lim}}=10^{((M_{\sun}-M_r)/2.5)}L_{\sun}$, where $M_{\sun}=4.52$ and $M_r=m_r-5\log(d_{\mathrm{lim}}/10))$. 
Thus, the sample luminosity
limits are $L>0.102\times 10^{10}h^{-2}L_{\sun}$, $L>0.256\times
10^{10}h^{-2}L_{\sun}$, $L>0.643\times 10^{10}h^{-2}L_{\sun}$ and $L>1.61\times 10^{10}h^{-2}L_{\sun}$.
In the catalogue by \cite{tago2010} the group luminosities are also corrected to include
the unobserved galaxies, but this correction is not necessary for volume-limited samples, and we have restored the uncorrected
values. The group catalogue is compiled using a version of the FOF-algorithm in which 
the linking length varies with the volume-limited galaxy
sample, but it is fixed inside a specific sample. These lengths are
0.250, 0.31, 0.41 and 0.54 $h^{-1}$Mpc for S1, S2, S3 and S4,
correspondingly (see more details in Tago et al. \citeyear{tago2010}).

The total numbers of groups in the different samples are 5463, 12590, 18973
and 9139 in S1, S2, S3 and S4, correspondingly. Roughly half of the
groups are actually galaxy pairs. To scale all the absolute
numbers to spatial densities that can be used for comparison, we have calculated the total volumes of the
different volume-limited samples. SDSS coverage calculated
  directly from the data used in this study is 7221 square
  degrees. Correspondingly, the volumes for different samples are 1.517$\times 10^6$, 6.541$\times 10^6$, 
24.27$\times 10^6$ and 83.78$\times 10^6$ $h^{-3}$Mpc$^{3}$ for S1, S2, S3 and S4.

We divide our groups in different classes on the basis of group
richness. 
Galaxy pairs are special groups that have only two members; so
we divide the groups into classes
with $N_{\mathrm{gal}}=$2, 3--9 or at least ten galaxies. The fractions of groups in the
volume-limited samples in each
class are given in Table ~\ref{tab1}. We notice that at least 60 percent
of all groups are galaxy pairs and the fraction of groups with at
least 10 members is only $<$2 percent. 
Roughly, $\sim$30 percent of groups are intermediate groups with $N_{\mathrm{gal}}=$3--9
members.
The fraction of pairs increases from 0.64 to 0.8 with the luminosity limit of a sample 
as faint galaxies are dropped.
We also give the absolute numbers of groups
in different samples in the table.

\begin{table}
 \caption{Fractions of galaxy groups in different richness classes in
   observations. 
The numbers of groups $N_{\mathrm{g}}$ in the Tago et al. (\citeyear{tago2010})
catalogue 
are given for illustration.} 
 \label{tab1}
 \centering
\begin{tabular}{cccc}
 \hline \hline\\[-8pt]
 Sample & pairs & $N_{\mathrm{gal}}=$3--9 & $N_{\mathrm{gal}}\ge 10$ \\[6pt]
 \hline
S1: M $\le -18$ & 0.64 &0.34  &0.021 \\
$N_{\mathrm{g}}$ & 3498 & 1965 & 117\\
S2: M $\le -19$ & 0.65 &0.33  &0.016 \\
$N_{\mathrm{g}}$ & 8242 & 4348 & 200\\
S3: M $\le -20$ & 0.68 &0.31  &0.010 \\
$N_{\mathrm{g}}$ &12829 &6144 & 196\\
S4: M $\le -21$ & 0.80 &0.20  &0.00066 \\
$N_{\mathrm{g}}$ & 7266 & 1873 & 6\\
\hline
\end{tabular}
\end{table}

Tago et al. (\citeyear{tago2010}) compiled also a group catalogue
based on the full magnitude-limited galaxy sample. 
This includes all galaxies that have the Petrosian
$r$-magnitudes 
between $r=12.5$ and $r=17.77$. 
The comparison between the volume-limited galaxy
groups and magnitude-limited galaxy groups can be used to quantify how the removed
galaxies in a certain volume-limited sample affect group
definition. For this reason we have calculated the fractions of galaxy
groups that have the same number of members in both samples (see Table \ref{tab2}).

\begin{table}
 \caption{Fractions of galaxy groups that have the same richness
   in the volume-limited and magnitude-limited SDSS group catalogues.}
 \label{tab2}
 \centering
\begin{tabular}{cccc}
 \hline \hline\\[-8pt]
 Sample & pairs & $N_{\mathrm{gal}}=$3--9 & $N_{\mathrm{gal}} \ge 10$ \\[6pt]
 \hline
S1: M $\le -18$ & 0.41 &0.20  &0.0085 \\
S2: M $\le -19$ & 0.42 &0.21  &0.015 \\
S3: M $\le -20$ & 0.44 &0.22  &0.010 \\
S4: M $\le -21$ & 0.46 &0.29  & - \\
\hline
\end{tabular}
\end{table}

More than 40 percent of all galaxy pairs are pairs in both
the volume-limited and magnitude-limited samples. For groups with $N_{\mathrm{gal}}=$3--9
members, 20--30 percent of all groups have the same richness.
It is also possible to calculate the fractions of galaxy groups in the
magnitude-limited catalogue that are separated into several groups in a
volume-limited catalogue. The fractions of multiple groups in a
volume-limited catalogue 
that are single groups in the magnitude-limited
catalogue are 0.11, 0.097, 0.071 and 0.025 in the S1, S2, S3 and S4, respectively.

This comparison between the magnitude-limited catalogue and the
volume-limited catalogues 
shows that in reality the group richness and galaxy
content are strongly dependent on the magnitude limits. Only a small
fraction of groups have exactly the same galaxies in both catalogues. 
This affects the comparison between the observed and simulated
  groups, but if the data is calculated in the same way for both groups, the
  comparison is still reliable.

\subsection{Simulations}
The Millennium Simulation (Springel et al. \citeyear{springel}, hereafter
MS) is a cosmological N-body simulation 
of the $\Lambda$CDM model performed by
Virgo Consortium, using a customized version of the GADGET2 code.
The MS follows the evolution of $2150^3$ particles from the redshift
$z=127$ in a box of 500 $h^{-1}$Mpc 
on a side. The cosmological parameters of the MS simulation are: $\Omega_{\mathrm{m}}=\Omega_{\mathrm{dm}}+\Omega_{\mathrm{b}}=0.25$, 
$\Omega_b=0.045$, $h=0.73$, $\Omega_{\Lambda}=0.75$, $n=1$, and
$\sigma_8=0.9$.
These values are slightly off from the current best estimate values
based on the Planck data:
$h=0.678$, $\Omega_{\mathrm{m}}=0.315$, $n=0.961$ and
  $\sigma_8=0.829$ (Planck Collaboration XVI 2013).
This may have a small effect on the differences that we see in our
study, but this effect is very difficult to quantify.
We choose the data from the $z=0$ snapshot, since the magnitudes in the
observational catalogue correspond to the $z=0$ magnitudes and possible
evolutionary effects are so small that we do not expect that they
influence the results.

In our analysis we use two different semi-analytic galaxy formation
models: 
Font et al. (2008) and Bertone et al. (2007) data. These are based on different 
galaxy formation models (GALFORM (Durham model) and L-Galaxy (Munich model), 
respectively. These differ 
from each other in the halo/subhalo merger tree schemes, as well as in 
details of the baryonic physics. Several authors 
(e.g. for GALFORM: Bower et al. 2006, Croton et al. 2006, 
Font et al. 2008, for L-Galaxy: De Lucia et al. 2006, 
De Lucia et al. 2007, Guo et al. 2011) have adjusted SAMs for 
both schemes to reach better agreement with the observational data. 
These analyses have shown, for example, that the amplitude of the 
galaxy luminosity function depends strongly on the 
feedback models. Especially, the bright end of the galaxy luminosity function 
can be reduced by active galactic nuclei (AGN) feedback models.
 
Based on the GALFORM scheme, Bower et al. (2006) introduced improved procedure 
for the feedback from the AGN and the 
growth of the supermassive black holes.
Moreover to improve agreement with satellite galaxy 
colour distributions with observations, Font et al. (2008) implemented 
into Bower et al. (2006) version the prescription
for ram pressure stripping of the hot gaseous haloes. 
Similarly, Bertone et al. (2007) developed previous 
SAM (De Lucia \& Blaizot et al. 2007) including the AGN 
feedback by Croton et al. (2006) and a new procedure for the SN feedback 
stellar wind model and for the dust attenuation.

The galaxies from the
semi-analytical procedure by Bertone, De Lucia and Thomas (2007) are
used to construct the main mock catalogue. 
The Bertone model predicts a lower number of dwarf galaxies than many other
models, a feature that fits better with observations.
The drawback of their model is that it
predicts a larger number of bright galaxies than found in observations.
The semi-analytical galaxy catalogue used includes all the necessary
information for direct comparison of galaxy group properties.
To build a group catalogue for the Bertone SAM galaxies we use the
  full simulation box that is a cube with the size of
500 $h^{-1}$Mpc.

To understand how sensitive our results are for different SAMs, we
used the SAM galaxy catalogue by Font et al. (2008).
For this sample we also used 
a cube with the size of 500 $h^{-1}$Mpc. 
The Font data is used as the reference sample, 
to show general trends of distributions for
qualitative evaluation. The Bertone SAM data is used for the main analysis.

To construct a simulated galaxy group catalogue we start from the
general collection of
SAM galaxies in the simulation.
In the first step, to mimic the observations the galaxies inside the simulation box are
transferred to the redshift space. 
Since the number of groups in observations that are influenced by
  the edge effects is so small, we don't include the actual SDSS
  footprint to our analysis. We expect that the error caused by this
  is so small that it can't be seen in our analysis.
In this process a coordinate transformation is applied, where one
corner of the cube is taken as the position of the observer (the coordinates
0, 0, 0) 
and each galaxy is shifted in the line-of-sight direction by its
speed vector projection in the line of sight. 
In this step we include 30 km $\mathrm{s}^{-1}$ rms error in redshift with Gaussian
  distribution. However, our tests show that this has only a
  marginal effect on the distributions and all the conclusions are the same even if this error is ignored.
Then the volume limit cuts, as stated earlier, are applied to the
data sets, but here we use different abbreviations for the samples:
M1, M2, M3 and M4.
For the M4 sample a smaller upper cut is applied ($z$ defined by the
distance of 462 $h^{-1}$Mpc due to the sample limit), 
because we want to avoid problems caused by sample
asymmetries. 
After that, we use precisely the same FOF-code, with the same linking lengths as
used in Tago et al. (\citeyear{tago2010}).
Hence, the method used to construct the SDSS groups and simulated groups is
  exactly the same and the remaining differences should be due to the 
spatial distribution of galaxies in the large-scale structure and/or in the
properties of galaxies. In Sec.~3.4 we will study, for
example, how very close galaxies that would very likely be missed in the SDSS data,
may cause some differences. However, since this incompleteness is internal
to the published DR7 galaxy catalogue, we cannot model this very accurately.

Using the calculated comoving volumes we normalize the galaxy
group property distributions and the spatial densities of simulated
data.
This allows direct comparison between the volume-limited samples in simulations and observations.
The spatial densities of all groups and the numbers of the groups $N_{\mathrm{g}}$ for different
luminosity limits and for group definitions are given in 
Table \ref{tab3}.
The number densities of groups in simulations and observations 
are similar (within the factor of
$\sim1.5$), but systematically slightly
larger for simulated groups in all samples compared with observations.
For large groups ($N_{\mathrm{gal}}\geq 10$) the difference is much larger
(by a factor of $\sim$2--6), depending on the sample.
The difference is probably due to SAM and the chosen
cosmology. Since the number density of galaxies in MS is larger, 
it is expected that by using the same linking length, there are more groups
in the simulated sample than in the SDSS sample. 
This can be seen by comparing the luminosity functions given in
Fig. ~\ref{halo_lum_function} shown later in the text.
Nevertheless, one should keep in mind that it is very
difficult to estimate the statistical errors in these values, since
the errors include the sample variance that cannot be estimated as we
have only one simulation. 

\begin{table}
 \caption{Numbers of galaxy groups with
   $N_{\mathrm{gal}}\geq 2$ and $N_{\mathrm{gal}}\geq 10$ 
    for the Bertone SAM galaxies.
The data is divided into four different samples. We give
    also the group number densities for the simulations (Bertone) and for
    the observations (Tago) in units of $h^{3}$Mpc$^{-3}$.
}
 \label{tab3}
 \centering
 \begin{tabular}{ccc}
 \hline \hline\\[-8pt]
Sample & $N\ge2$ & $N\ge10$ \\[6pt]
 \hline
M1: M $\le -18$ & 4419 & 209\\
Bertone $N_{\rho}$ & 4.02$\times 10^{-3}$ & 1.90$\times 10^{-4}$\\
Tago $N_{\rho}$ & 3.47$\times 10^{-3}$ & 7.43$\times 10^{-5}$\\
\hline
M2: M $\le -19$ & 9989 & 329\\
Bertone $N_{\rho}$ &2.04$\times 10^{-3}$  & 6.74$\times 10^{-5}$\\
Tago $N_{\rho}$ & 1.85$\times 10^{-3}$ & 2.94$\times 10^{-5}$\\
\hline
M3: M $\le -20$ & 21189 & 418 \\
Bertone $N_{\rho}$ &1.12$\times 10^{-3}$ &2.25$\times 10^{-5}$\\
Tago $N_{\rho}$ & 7.52$\times 10^{-4}$ & 7.77$\times 10^{-6}$\\
\hline
M4: M $\le -21$ & 15755 & 97\\
Bertone $N_{\rho}$ & 2.36$\times 10^{-4}$&1.45$\times 10^{-6}$\\
Tago $N_{\rho}$ & 1.05$\times 10^{-4}$ & 6.89$\times10^{-8}$\\
\hline
\end{tabular}
\end{table}

In Table \ref{tab4} we also give the median
and mean values for the group masses in the simulation. 
The masses given in the table include only the mass of the
  galaxies within the sample magnitude limits; faint galaxies
outside the observational window are not included. 
The group mass is simply the sum of all virial masses of the
  galaxies in the group.
This table describes the
typical group or clusters masses in different samples.
Groups with $N_{\mathrm{gal}}\ge 10$ members are an order of magnitude
more massive than the mean value for all groups.
Groups in M1 have masses $\sim$4--40$\times 10^{12}h^{-1}M_{\sun}$
and 
for the most distant groups in M4 the masses are considerably larger, between  
$\sim$0.5--2$\times 10^{14}h^{-1}M_{\sun}$.

\begin{table}
 \caption{Mean and median masses (in units of $10^{12}h^{-1}M_{\sun}$)
   of the galaxy groups in the Bertone SAM data.}
 \label{tab4}
 \centering
 \begin{tabular}{ccc}
 \hline \hline\\[-8pt]
Sample & Mean & Median \\[6pt]
 \hline
M $\le -18$, All& 4.22$\pm 0.3$ & 1.10 \\
Pairs& 1.15$\pm 0.04$ & 6.49\\
3--9& 4.35$\pm 0.2$ & 2.37\\
$\ge10$& 42.2$\pm 5$ & 22.1\\
\hline
M $\le -19$, All& 6.18$\pm 0.2$ & 1.73\\
Pairs& 2.04$\pm 0.04$ & 1.11 \\
3--9& 7.61$\pm 0.2$ & 4.04\\
$\ge10$&  66.8$\pm 5$ & 37.6\\
\hline
M $\le -20$, All& 13.7$\pm 0.3$ & 4.25\\
Pairs& 5.93$\pm 0.1$& 2.82\\
3--9& 22.4$\pm 0.6$& 10.6\\
$\ge10$& 163$\pm 10$& 101\\
\hline
M $\le -21$, All& 46.7$\pm 1$& 23.2\\
Pairs&  30.1$\pm 0.8$ &17.6\\
3--9&   97.7$\pm 4$& 58.9\\
$\ge10$& 238$\pm 40$ & 235 \\
\hline
\end{tabular}
\end{table}

\section{Comparison between galaxies in the simulation and the SDSS galaxies}
Before we start to compare the group properties of simulations
with observations, we analyse the galaxy luminosity functions.
We give the galaxy number densities in different volume-limited samples
in Table \ref{tab5}.

\begin{table}
 \caption{Galaxy number densities in units of $h^{3}$Mpc$^{-3}$ in the Millennium simulation (MS) and SDSS.}
 \label{tab5}
 \centering
\begin{tabular}{cccc}
 \hline \hline\\[-8pt]
 Sample & Bertone & Font & SDSS \\[6pt]
 \hline
M $\le -18$ & 3.26$\times 10^{-2}$ & 4.00$\times 10^{-2}$ & 2.58$\times 10^{-2}$\\
M $\le -19$ & 1.92$\times 10^{-2}$ & 2.12$\times 10^{-2}$ & 1.40$\times 10^{-2}$\\
M $\le -20$ & 8.26$\times 10^{-3}$ & 7.78$\times 10^{-3}$ & 5.88$\times 10^{-3}$\\
M $\le -21$ & 1.66$\times 10^{-3}$ & 1.88$\times 10^{-3}$ & 1.20$\times 10^{-3}$\\
\hline
\end{tabular}
\end{table}

This Table shows certain differences between simulations and observations.
Firstly, all cosmological simulations are realizations
of a set of initial conditions and the subsequent large-scale structure
is always different. Secondly, all SAMs
populate haloes with galaxies in a different way and include
different physical processes and approximations
in their recipes.

\begin{figure}
 \centering
 \resizebox{\hsize}{!}{\includegraphics*[angle=0]{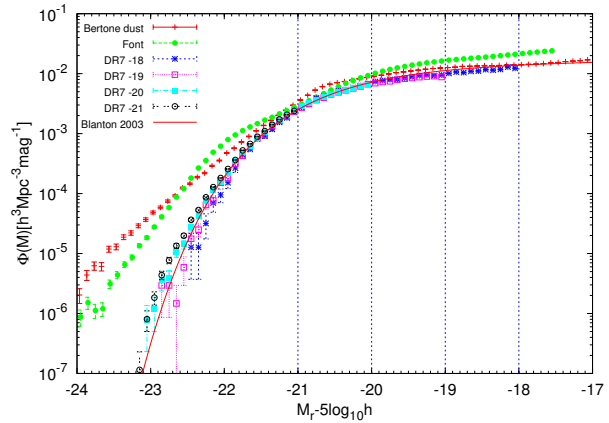}}
 \caption{The luminosity function for the $r$-band of all galaxies in the
   Bertone SAM based on the Millennium Simulation (MS),
   compared with the observed SDSS luminosity
   function as given in \protect\cite{tago2010}, for four different volume-limited samples.
The luminosity function derived in Blanton et al. (2003) is given for comparison.
}
\label{halo_lum_function}
\end{figure}

The galaxy luminosity functions for SDSS and for the \cite{bertone} and \cite{font} SAM galaxies 
applied to the Millennium
data, are shown in  Fig. ~\ref{halo_lum_function}.
For the Bertone data we use dust corrected magnitudes, but for the Font data
only $M_r$ is given. 
This may introduce a small shift in the luminosity function for the Font
data (see, e.g., Tempel et al. (2011)). Both SAMs include AGN feedback.
In the same figure we show the luminosity limits used in our volume-limited samples as vertical lines. 
The agreement between the
observations and simulations is relatively good between
$-21.5<M_r-5\log h<-18$, 
but for the magnitudes $M_r-5\log h<-21.5$ the Millennium SAM data clearly overestimate the number
of luminous galaxies. 
\cite{font} compared the simulations against the 2dFRS data and
  they agree rather well, but our study shows that their SAM data do not
  agree that well with the SDSS data. The most feasible reason for this is in intrinsic differences of
  the SDSS and 2dFGRS galaxy luminosities, since these galaxy surveys
  have different sky coverage and wavebands. Differences of the
  surveys are seen especially at the bright end of the luminosity
  function.

It seems that this problem is very difficult to eliminate
(especially for the $r$-band) in
the SAM procedure; the same problem remains in a recent study by
Guo et al. (2011), albeit not that severe.
Although the difference seems to be quite
drastic, 
for our analysis this difference is not that significant, 
since most galaxies belong to the region where all distributions agree
quite well.
The most luminous galaxies are typically in rich groups that consist of rare
objects, and therefore in the statistical study their effect on the
general group properties is rather small.

In the Fig. 1 we also show the luminosity function calculated by
\cite{blanton}. They derived the best-fit Schechter function
using the older version of the SDSS catalogue and obtained the values for
the parameters 
$\Phi_{\ast}=(1.49\pm0.15)\times 10^{-2}h^3 \mathrm{Mpc}^{-3}$, 
$M_{\ast}-5\log_{10}h=-20.44\pm0.01$ and 
$\alpha=-1.05\pm0.01$. 
This agrees very well with the \cite{tago2010} galaxy
data.

\subsection{Group luminosity functions}
\begin{figure*}
 \centering
 \resizebox{\hsize}{!}{\includegraphics*[angle=0]{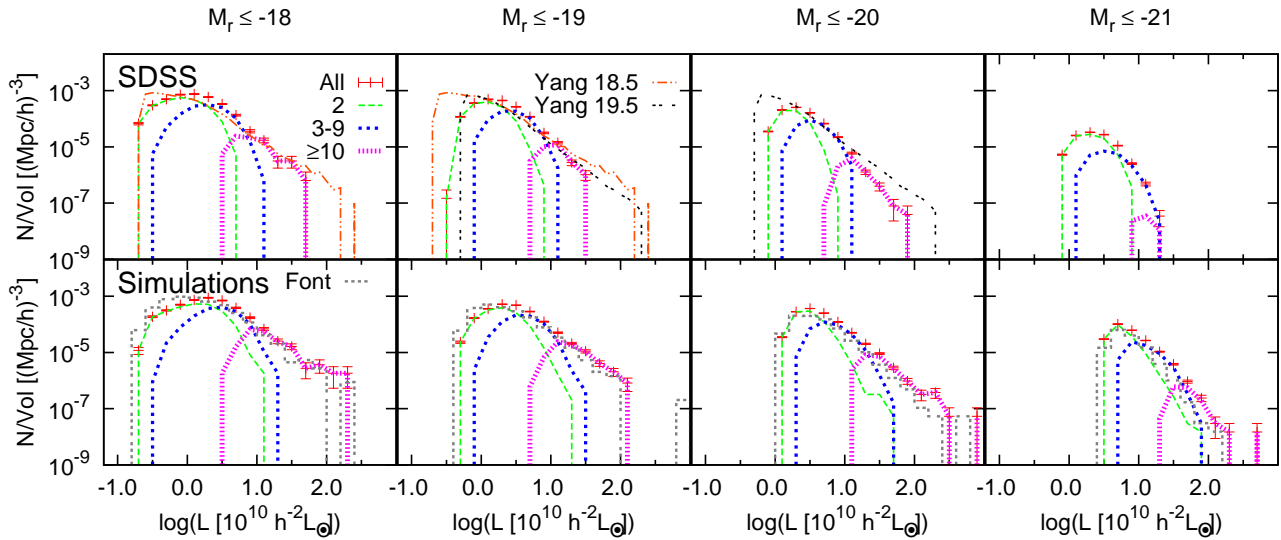}}
 \caption{$R$-band luminosity functions for groups in the SDSS data
    (upper row) and
   for the group catalogue obtained from the Bertone SAM based on MS (lower row), for four different volume-limited samples. 
All groups with two or more member galaxies are included. 
In all panels we show the total distribution as red points and 
the Poisson errors in the bins by red bars. The distributions for galaxy pairs
(green dashed line), for groups with $N_{\mathrm{gal}}=$3--9 members (blue dotted line) and for
groups with $N_{\mathrm{gal}}$$\ge 10$ members (magenta dash dotted line) are also given.
The luminosity functions for the DR4 groups
   by \protect\cite{yang2009} are shown as two dashed
   lines for two different samples.
}
\label{group_luminosity}
\end{figure*}

Here we compare the group luminosity functions that give the
  number of groups per unit volume as a function of total group
  luminosity ($L_{\mathrm{gr}}$). The luminosities of all 
galaxies that belong to the group are summed together to calculate
the group luminosity $L_{\mathrm{gr}}=L_{\mathrm{main}}+\sum{L_{\mathrm{satellite}}}$, where 
$L_{\mathrm{main}}$ is the luminosity of the main galaxy and $L_{\mathrm{satellite}}$ is
the luminosity of a satellite galaxy in the same group. 
The main galaxy is always the most massive galaxy in the group
  obtained from the simulation and in the observations, the main galaxy is the brightest galaxy in a group.
The luminosity functions for each volume-limited sample are given in Fig.~\ref{group_luminosity}. 
The lower limit of the group luminosities 
is determined by the luminosity limit for individual galaxies and therefore the distribution gets 
more narrow for bright groups. In observations there are
systematically less bright groups 
than in simulations, but the overall agreement is quite good. 
This difference is mostly because in simulations there are more luminous galaxies.
The agreement is very good in the $M \le-18$ sample for intermediate
groups ($N_{\mathrm{g}}=$3--9), and the greatest differences are seen for rich
groups.
The difference is quite drastic in the $M \le-21$ sample, for which the
whole distribution is shifted towards smaller magnitudes.

The fact that simulations give 
too many and too bright groups is partly due to too high luminosities of some
galaxies (see the galaxy luminosity function). Our more detailed analysis of the
Bertone SAM data revealed that there are some galaxies with
exceptionally low $M/L$ ratio. For this reason the bright end of the
luminosity function differs considerably from the observations. In the dust-corrected
data, the number of these luminous and light galaxies is reduced, but
obviously not all of them. About 7--10 percent of Bertone's dust-corrected galaxies at $M \le -22$ seem still to have exceptionally low
$M/L$ ratio at the given mass. Extracting these objects from Bertone's
data gives a luminosity function which fits better with observations. This modification
changes mainly the bright end of the luminosity function, which would not be the result
of the variation of the $\sigma_8$ parameter. This hints that dust
correction is too inefficient
 in SAMs or there are problems modelling the physics
of galaxy formation.
Guo et al. (2011) compared the SDSS luminosity function with 
their SAM and they reported, that their model overpredicts the 
abundance of luminous galaxies likely because of problems in their dust modelling.  
This does not exclude the effect,
which may come from the possible difference of $\sigma_8$ between
theory and observations. For example, Yang et al. (2005) showed that 
the multiplicity and luminosity functions of groups are inconsistent
with the observational data if $\sigma_8=0.9$ is used, but
$\sigma_8=0.7$ gives a perfect fit. Also, van den Bosch et al. (2005)
showed that simulations with $\sigma_8=0.9$ cosmologies are unable to
match the abundances of central and satellite galaxies with
observations.

Another reason for the mismatch between simulations and observations is that
the richness distribution is different for rich groups (see Fig. 3). In simulations
there are more rich groups that increase the number of luminous groups.
Especially, the differences are notable for rich groups in the last
two volume-limited samples. 
These samples typically contain only the most
luminous galaxies and it is expected that the group
luminosity is also overestimated, due to the unrealistically bright
massive galaxies.
The Font data and the Bertone data MS
both give
almost identical distributions for all samples. 
Hence, different SAMs give very similar results although the
luminosities of galaxies are modelled in different ways.

To compare our group luminosity functions to other observational studies we use the
data from the study by \cite{yang2007}. They
constructed the group catalogue based on the DR4 data using an adaptive halo based
group finder (Yang et al. 2005b) and calculated
the corresponding volume-limited samples and the luminosity functions
for the groups $\Phi(L_{gr})$. They divided the sample into two different
classes: groups with $M_r-5\log h \leq -19.5$ and groups with
$M_r-5\log h \leq -18.5$. 
Since our magnitude limits differ slightly from theirs
and direct comparison is not possible, we show both
distributions in our figures. The correspondence is relatively good,
although the method for group construction is
different. 
The main difference is at large luminosities. The Tago et
al. (2010) group catalogue does not contain as many luminous groups 
($\log{L_{gr} [10^{10}L_{\sun}]} \ga 1.8$)
as the \cite{yang2009} catalogue. In this respect the simulations
agree better with \cite{yang2009} data than with the \cite{tago2010} catalogue
although simulated groups have been constructed using the same algorithm
as in \cite{tago2010}.

\subsection{Group richness}
\begin{figure*}
 \centering
 \resizebox{\hsize}{!}{\includegraphics*[angle=0]{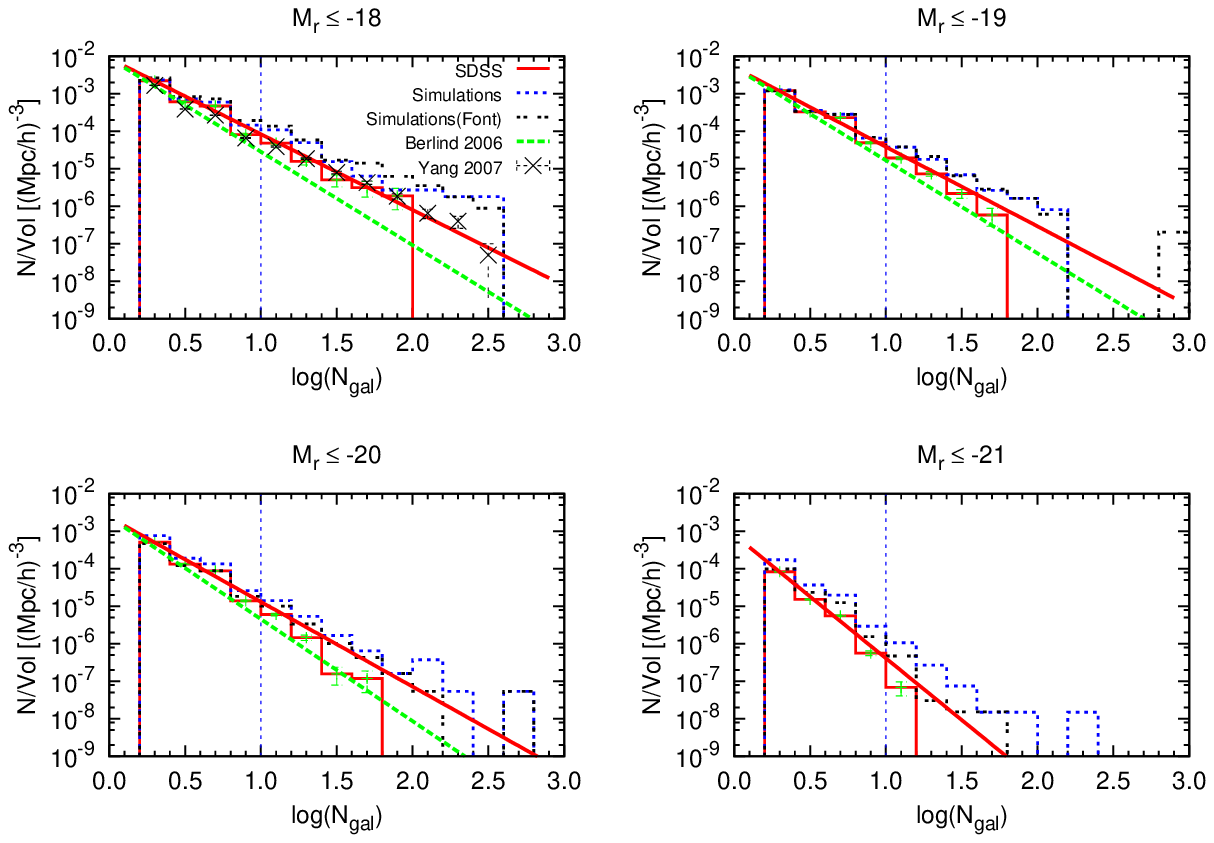}}
 \caption{Comparison of the group richness (number of galaxies in a group) in the
   SDSS (solid red line histogram) 
and in the Bertone SAM based on MS.
The Font data is given for comparison as well as the slopes given in 
Berlind et al. (2006) for the first three samples. Best-fitted
  slopes for the SDSS data are shown as red solid lines.
In the first panel the richness
distribution by Yang et al. (2007) is also shown.
}
\label{richness}
\end{figure*}

Multiplicity function 
estimates provide one of the key constraints on the galaxy group properties. 
The group richness distributions
  $N_{\mathrm{gal}}=N_{\mathrm{satellite}}+1$ 
(the number of galaxies in a group, here $N_{\mathrm{satellite}}$ is the number of satellite
galaxies), 
are shown in Fig. ~\ref{richness}. The first bin is for pairs and the distributions extend up to 
100 members. There is a clear exponential trend and we fitted straight lines to the
distributions to see how the slope values depend on the absolute magnitudes. 
For the SDSS groups \cite{berlind} found that the multiplicity functions are well fitted by power-law 
relations, with the best-fit slopes of $-2.49 \pm 0.28$, $-2.48 \pm 0.14$ and $-2.72 \pm 0.16$ 
for the absolute magnitude limits of $M_r\le -18$, $-19$ and $-20$. 
These absolute magnitude limits are approximately the same as for our first three samples. 
The least-square fit to our observational data gives the values 
$-2.02 \pm 0.18$, $-2.12 \pm 0.17$, $-2.26 \pm 0.19$ and $-3.29 \pm 0.24$ for the
$M_r\le -18$, $-19$, $-20$ and $-21$ samples, respectively. 
To include the Poisson errors we used $1/\sqrt{N}$ weights during the fit, where $N$ is the number of
groups in a bin.  
Our values are slightly smaller from those of \cite{berlind}, but there is a similar 
trend that for brighter groups the slope is steeper. 
Without using weights in the fit, the slope values are closer
to the ones obtained by \cite{berlind}.
Since our samples in simulations consist of groups with different mean masses (see
Table 4), we can conclude that richness distribution is a function of
group mass. For massive groups the distribution is steeper than for the
less massive groups. Analogously, the same should apply for
observational data.

For small groups (the first three bins) the agreement between all groups is very good. 
As the number of members in the groups starts to exceed 10, the distributions start
to deviate.  The MS group richness distribution, that 
should be the same in the ideal case, is above 
the SDSS distribution. As for luminosity
distributions, the Font data and Bertone
data give results that are very close to each other. 
Thus, we see that in the Millennium SAM data there are too many rich and luminous
groups compared with observations and the differences are larger for richer groups.

The richness distribution can also be calculated for the \cite{yang2007}
group catalogue. We used their sample III that
contains 300049 groups extracted from the SDSS DR4 galaxy data. Their
catalogue of galaxy groups is not volume-limited, but it is still
interesting to compare the richness distribution. They used the magnitude limit 
$M_r-5\log{h}\leq -19.5$ that is between these for our two first samples. 
The richness distribution is shown in the first
panel (crosses) with an arbitrary scaling, since the total volume of their
catalogue is not known. The agreement between our SDSS richness
function and theirs is very good. 

It is also interesting to compare our results with those for the more recent GAMA galaxy group
catalogue (Robotham et al. \citeyear{robotham}). Their group
catalogue is not volume-limited either, but the comparison is still valuable.
Their richness distribution also follows a linear trend, especially
for their mock catalogue, with the estimated slope value of $\sim -1.7$.
It is expected that the distribution is not as steep as in the volume-limited catalogue, since in their magnitude-limited sample there are more
faint galaxies and group richnesses are typically higher.
They noticed that the number of high-richness systems is
significantly 
different between the observational data and the
mocks, but for low-richness systems the distributions agree better. 
This is exactly the same feature that we see in our data.
Their interpretation is that SAM (Bower et
al. 2006, GALFORM) populates more massive haloes with an excessive number of faint satellite
galaxies, a feature that is reported to be a problem in this model (Kim et al. 2009).

\subsection{Projected virial radii estimator}
\begin{figure*}
 \centering
 \resizebox{\hsize}{!}{\includegraphics*[angle=0]{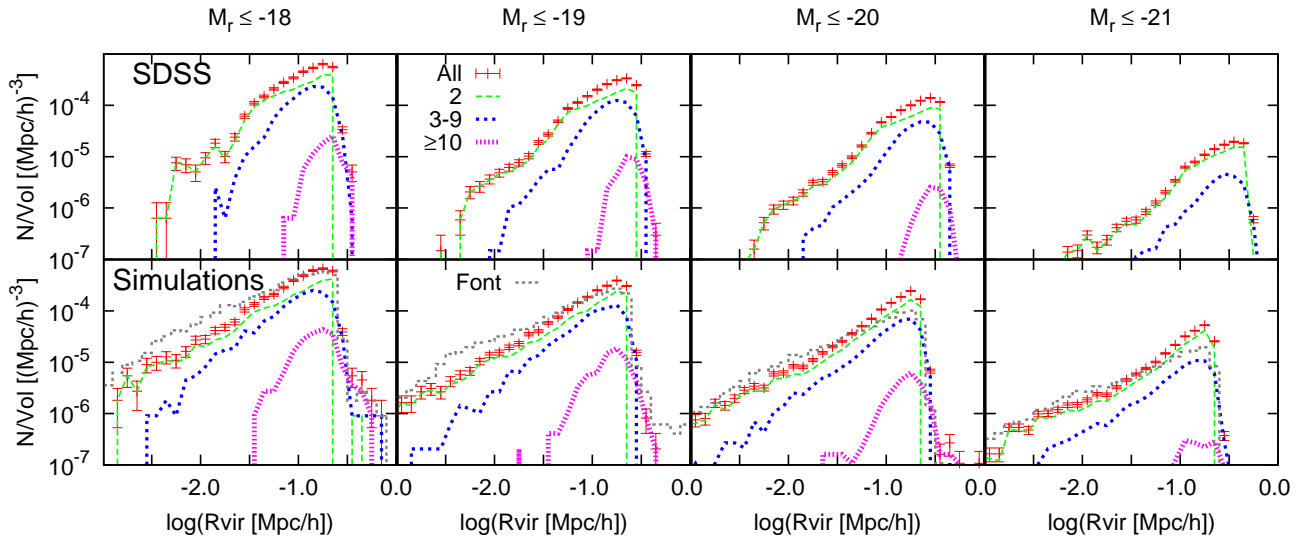}}
 \caption{Comparison of the distributions of virial radii for SDSS
and for Bertone SAM based on MS. The total distribution is shown
  as red points with
the Poisson errors in the bins. The distributions for galaxy pairs
(green dashed line), groups with $N_{\mathrm{gal}}=$3--9 members (blue
dotted line) and
groups with $N_{\mathrm{gal}}$$\ge 10$ members (magenta dotted line) are also given.}
\label{virialradius}
\end{figure*}

One of the basic parameters of a galaxy group is its virial radius.
To estimate the projected virial radius we use the harmonic mean of the projected separations:
\begin{equation} 
R_{\mathrm{vir}} = \left(\frac{1}{n_{\mathrm{pairs}}}\sum{\frac{1}{r_{ij}}}\right)^{-1} ~,
\end{equation} 
where $r_{ij}$ is the mutual projected angular diameter distance between galaxies $i$
and $j$ for $n_{\mathrm{pairs}}$ pairs. If the number of galaxies in a group is
$n$, then $n_{\mathrm{pairs}}=n(n-1)/2$. For large groups this estimator
  becomes tightly correlated with the real virial radius, but this equation is not good for small
groups. However, it is used here, since in the SDSS comparison paper it was
calculated in this way and we call it simply as virial radius.
We will use the virial radius estimator as a scaling factor in the analysis of radial
distance distribution of galaxies in a later section.
In the simulation we know also the virial radius of the main galaxy, but this and the virial radius of the group
are only weakly correlated. 
The distributions of projected virial radii for all groups are shown in Fig. ~\ref{virialradius}. 

In Table \ref{tab6} we show the mean values of the virial radii for simulations and observations.
For these estimates we have ignored galaxy pairs, since these bias the results
towards small values.
For the observations, the mean values vary between 0.14 and 0.36 $h^{-1}$Mpc  
and a general trend is that $\langle R_{\mathrm{vir}} \rangle$ is larger for those
  groups that do not include faint galaxies (M $\le -21$) compared
  with the groups that include also faint galaxies (M $\le -18$).
Rich groups are also typically larger in size than loose
groups and therefore $\langle R_{\mathrm{vir}} \rangle$ is systematically larger for rich groups.
The sample M $\le -21$ extends much further in distance than the M $\le -18$ sample
and the linking length in the FOF-algorithm increases with the absolute magnitude limit.
Therefore there are much fewer faint galaxies (less massive) in the M
$\le -21$ sample than in the  M $\le -18$ sample and at large distances there
are fewer small groups that have only a few members, if all
  galaxies could be observed. 
In the simulations $\langle R_{\mathrm{vir}} \rangle$ does not behave in the same way --
it is close to 0.12--0.16 $h^{-1}$Mpc for all the samples.

\begin{table}
 \caption{Mean values of the virial radii for all $N_{\mathrm{gal}}=$3--9
   and $N_{\mathrm{gal}}$$\ge 10$ distributions shown in Fig. ~\ref{virialradius}, in
   $h^{-1}$Mpc units.}
 \label{tab6}
 \centering
 \begin{tabular}{ccc}
 \hline \hline\\[-8pt]
Sample, $N_{\mathrm{{gal}}}$ & Simulations
 (Mean) & SDSS (Mean) \\[6pt]
 \hline
M $\le -18$, 3--9 & 0.122$\pm 0.002$ & 0.136$\pm 0.001$  \\
$\ge10$ & 0.165$\pm 0.005$ & 0.200$\pm 0.005$ \\
\hline
M $\le -19$, 3--9 & 0.123$\pm 0.001$ & 0.165$\pm 0.001$ \\
$\ge10$& 0.164$\pm 0.003$ & 0.242$\pm 0.004$ \\
\hline
M $\le -20$, 3--9 & 0.128$\pm 0.0008$ & 0.216$\pm 0.001$ \\
$\ge10$ & 0.166$\pm 0.003$ & 0.307$\pm 0.006$ \\
\hline
M $\le -21$, 3--9  & 0.120$\pm 0.001$ & 0.274$\pm 0.003$ \\
$\ge10$& 0.156$\pm 0.007$ & 0.355$\pm 0.03$ \\
\hline
\end{tabular}
\end{table}

For all distributions, there are clear differences between the simulations
and the SDSS data (Fig. ~\ref{virialradius}). 
Firstly, in the simulations there are many more small groups with $\log R_{\mathrm{vir}}<-1.5$
than in observations. The agreement is the
best near the mode, but before and after the differences are large. For loose and rich groups
the distributions are shifted towards larger $R_{\mathrm{vir}}$-values as we
move from the M $\le -18$ sample to the M $\le -21$ sample. 
Numerically the effect of galaxy pairs on the distributions is remarkable.
The distributions are suddenly cut at a certain distance
due to the linking length value in the FOF-algorithm.
The observational limitations that explain the discrepancy between the SDSS and the
mock data at small $R_{\mathrm{vir}}$-values will be described in more detail in
the next section where we study the maximum projected size distributions.
Fig.~\ref{virialradius} shows that if 
galaxy pairs are removed from the analysis, the agreement between different samples is much
better and the shapes of the distributions are closer to each other.
However, the same observational bias affects also loose and rich groups, but not that significantly.
The Font data and the Bertone data give both rather similar distributions.

\subsection{Group sizes}
\begin{figure*}
 \centering
 \resizebox{\hsize}{!}{\includegraphics*[angle=0]{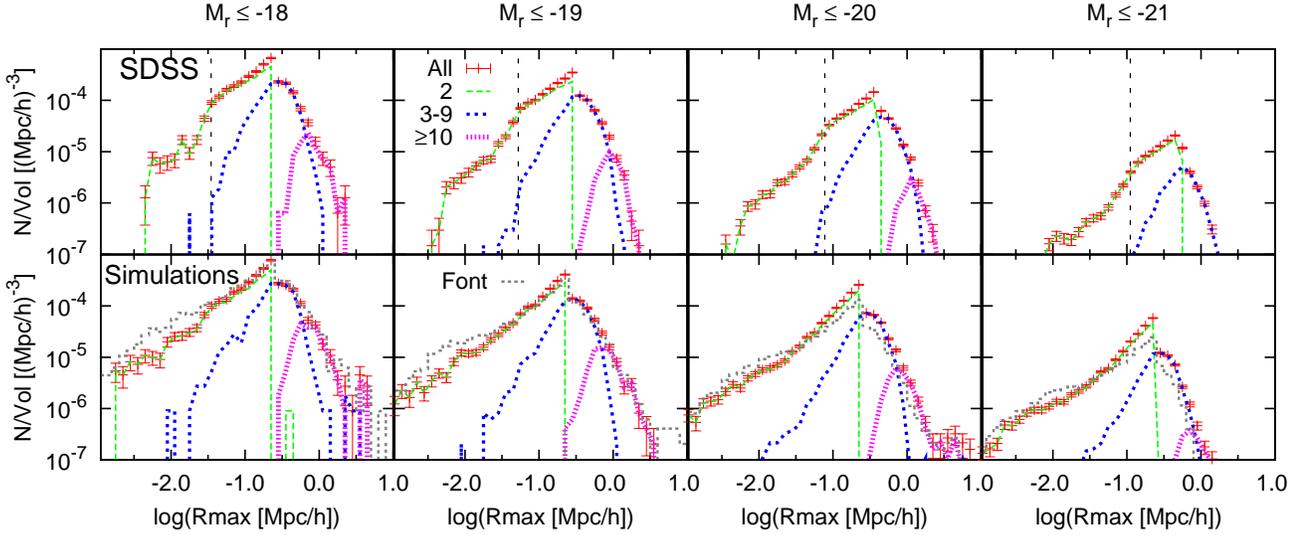}}
 \caption{Comparison of the group size distributions for SDSS and
   for Bertone SAM based on MS. The total distribution is shown
  as red points together with
the Poisson errors in the bins. The distributions for galaxy pairs
(green dashed line), groups with $N_{\mathrm{gal}}=$3--9 members (blue dotted line) and
groups with $N_{\mathrm{gal}}$$\ge 10$ members (magenta dotted line) are also
given. Black dashed vertical lines
   in the SDSS panels show the incompleteness limits calculated using
   the 55$\arcsec$ criterion described in chapter 3.4.}
\label{maxsize}
\end{figure*}

Another parameter that can be used to characterize a group is  
the maximum projected galaxy pair separation $R_{\mathrm{max}}$ in the group. We take that as the definition of the group size. 
The group size distributions are very similar to the $R_{\mathrm{vir}}$ distributions (Fig. ~\ref{maxsize}), but 
the disagreement between the simulations and observations is more
obvious for $\log(R_{\mathrm{max}}/h^{-1}\mathrm{Mpc})<-1$ . 
The largest differences are again for small groups. 
For the first two samples, the simulations have rather similar distributions to the SDSS
distributions for large groups, excluding galaxy pairs.
The group size distributions have mode values at $\sim 0.3$ $h^{-1}$Mpc, that is mostly due to the galaxy
pairs that peak at this distance. The peak is artificially sharp due
to the chosen linking lengths that vary between
0.25--0.54$h^{-1}$Mpc; the peaks are located close to the linking lengths.
We give in 
Table \ref{tab7} the mean values of the group sizes.

\begin{table}
 \caption{Mean values of group sizes for all $N_{\mathrm{gal}}=$3--9
   and $N_{\mathrm{gal}}$$\ge 10$ distributions shown in Fig. ~\ref{maxsize} (in
   $h^{-1}$Mpc units).}
 \label{tab7}
 \centering
 \begin{tabular}{ccc}
 \hline \hline\\[-8pt]
Sample, $N_{\mathrm{gal}}$ & Simulations
 (Mean) & SDSS (Mean) \\[6pt]
 \hline
M $\le -18$, 3--9& 0.290$\pm 0.005$ & 0.286$\pm 0.003$ \\
$\ge10$& 1.11$\pm 0.2$ & 0.788$\pm 0.03$ \\
\hline
M $\le -19$, 3--9& 0.293$\pm 0.002$ & 0.354$\pm 0.003$ \\
$\ge10$& 0.842$\pm 0.02$ & 0.948$\pm 0.02$ \\
\hline
M $\le -20$, 3--9 & 0.306$\pm 0.002$ & 0.467$\pm 0.003$ \\
$\ge10$& 0.954$\pm 0.04$ & 1.19$\pm 0.03$ \\
\hline
M $\le -21$, 3--9& 0.299$\pm 0.005$ & 0.576$\pm 0.006$ \\
$\ge10$& 1.18$\pm 0.15$ & 1.50$\pm 0.2$ \\
\hline
\end{tabular}
\end{table}
As for the $R_{\mathrm{vir}}$-distributions, for rich groups ($N_{\mathrm{gal}} \ge
  10$) the agreement between
mock groups and observations is relatively good. 
In observations the mean group sizes are systematically larger for the
M $\le -21$ than for  the M $\le -18$ sample. $\langle R_{\mathrm{max}} \rangle$ varies from
0.29$h^{-1}$Mpc to 0.58$h^{-1}$Mpc for loose groups and from
0.8$h^{-1}$Mpc to 1.5$h^{-1}$Mpc for rich groups. 
These values are $\sim 2$ times larger than for the virial radii.
In simulations the mean group sizes behave differently and they remain rather
constant (0.3$h^{-1}$Mpc) for the groups with $N_{\mathrm{gal}}=$3--9
for all samples. 
Only for the M $\le -18$ sample do the values agree quite well between
simulations and observations.
For groups with $N_{\mathrm{gal}}$$\ge 10$ the mean sizes are larger
for the M $\le -21$ sample than for the M $\le -18$ sample, but still
those are smaller than for the SDSS data, excluding the M $\le -18$ sample.
Also for the $R_{\mathrm{max}}$-distribution, the Font data produces similar results to
the Bertone data, although some differences can be seen.

The smallest groups, galaxy pairs, have been extensively studied by
\cite{patton} for the SDSS DR7 (for DR4 see also
\cite{ellison2010} and \cite{ellison2008}).
They found that the fraction of red galaxies in pairs is higher than that of a control
sample and the difference is likely due
to the fact that galaxy pairs reside in higher density environments than non-paired galaxies.
They also noted an important selection effect that needs to be
taken into account.
Fibre collisions in the original SDSS data cause small-scale
spectroscopic incompleteness 
at small galaxy separations (Strauss et al. \citeyear{strauss2002}).
In \cite{ellison2008} it is estimated that 
67.5 percent of pairs at angular separations below 55$\arcsec$ have been
missed due to this effect, even when the influence of 
overlap between adjacent plates and the use of two or more plates
in some regions is considered.
Taking into account the adopted cosmology and the 55$\arcsec$
criterion we have estimated the group size thresholds 
after which the incompleteness in the distributions at a
given distance starts (the smaller the size, the larger the
incompleteness). 
These limits for $\log(R_{\mathrm{max}} [h^{-1}\mathrm{Mpc}])$ are $-1.46$,
$-1.28$, $-1.11$ and $-0.955$ for different samples. 
In  Fig. ~\ref{maxsize} the limits are
shown as vertical dashed lines in the top panels.
Due to this selection effect, the group size distributions for the SDSS start
to deviate from the mock catalogue distributions near these limits.
It is expected that the same limitations affect also richer
groups and, in fact, this is also seen in Fig. ~\ref{maxsize}. 
The fraction of small groups with $N_{\mathrm{gal}}=$3--9 members in the SDSS is smaller than in simulations.
For $N_{\mathrm{gal}}$$\ge 10$ groups this effect is not that significant any more,
since the group sizes are typically very large. 
The observed distributions
are not as smooth as for the simulations.
The minimum size of the SDSS groups $\sim$10 $h^{-1}$kpc is also due 
to the same limitation.

\begin{figure*}
 \centering
 \resizebox{\hsize}{!}{\includegraphics*[angle=0]{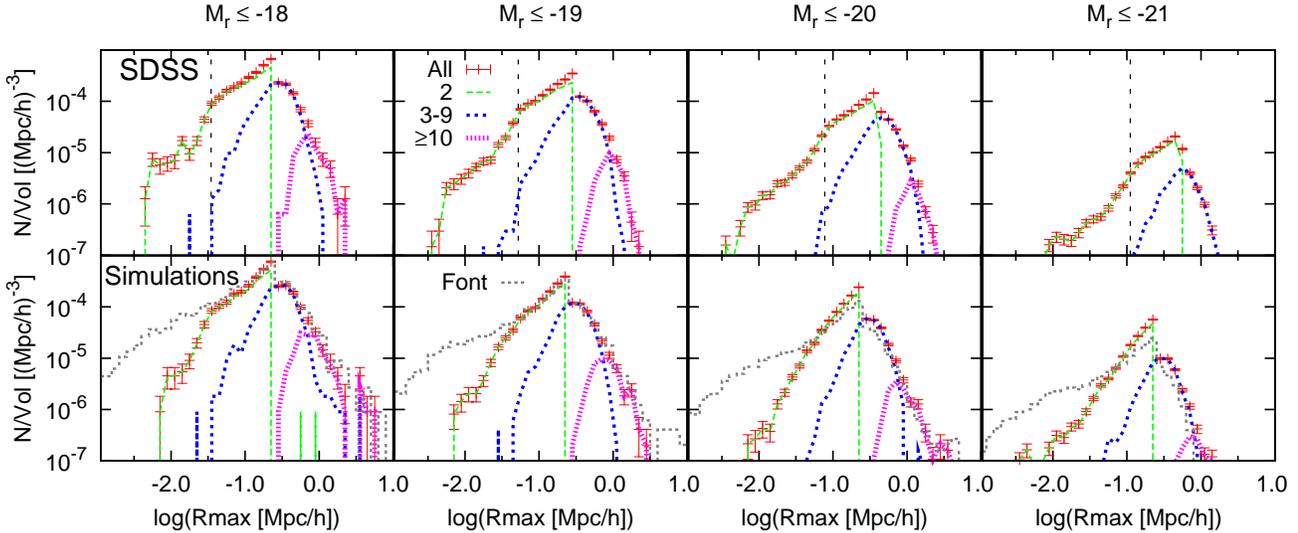}}
 \caption{Comparison of the group size distributions for the SDSS and
   for the modified mock data. All neighbour galaxies that
   lie closer than $< 55\arcsec$ are randomly removed from the
   simulated data. Otherwise the figure is the same as Fig. 5.}
\label{maxsize_nopairs}
\end{figure*}

To qualitatively test how close pair incompleteness could maximally
influence the results we removed randomly one of the galaxies, 
from all galaxies that are within
55$\arcsec$ (in projected separation) from each other in the mock data. 
By removing all close galaxies, we removed a number
of close pairs, but richer groups were also affected by this procedure 
(Fig. ~\ref{maxsize_nopairs}).
The size distributions agree then
much better with observations, especially for the $M \le -18$ and $M \le
-19$ samples. The distributions for larger groups agree also better, but the difference
is not as large as for pairs.
In the last two samples ($M \le -20$ and $M \le
-21$) there are significant differences as before, but the pair removal
behaves as expected. 
It is not surprising that this correction procedure does not give a
perfect match, since the 55$\arcsec$ rule is not absolute, and it is very difficult to model the incompleteness
effect precisely. 

Some of the differences in the distributions at small scales 
may also be due to the SAM and the used cosmology.
In the Bertone et al. (2007) model, subhalo orbits were 
followed under the influence of the tidal truncation and stripping 
until the resolution limit of the simulation was reached.  
Due to the limited resolution of the simulations, part of those 
subhaloes for which DM haloes are considerably disrupted, lose
their DM halo, but retain their galaxy properties formed earlier in the subhalo.
Once the resolution limit is reached, these 
''orphan" galaxies are placed on most bound DM particles of the former DM haloes. 
Subsequently, merger time (due to dynamical friction) of the orphan galaxy with the central galaxy is estimated 
using the analytical formula of Binney \& Tremaine (1987).
In Bertone et al. (2007), SAM orphan galaxies are not disturbed during 
the dynamical friction time. This assumption may overestimate orphan 
galaxy luminosities (Bryan et al. 2008). 
However, the orphans are not those galaxies that have exceptionally
low $M/L$ ratio mentioned in chapter 3.1.
Font et al. (2008) did not follow full orbital evolution of subhaloes and 
in their model subhaloes are merged according to the dynamical friction time formula. 
Thus, the Font et al. (2008) model does not include orphans.
In the Bertone et al. (2007) simulations 
orphans are denoted as "type2" haloes and we have included 
these in our analysis.

Weinmann et al. (2010) tested how the resolution 
limit affects the capability of MS
 (using the De Lucia \& Blaizot et al. (2007) model) 
to follow subhalo evolution.  
They concluded that MS
can be used to follow subhalo evolution to high 
accuracy for infalling subhaloes with 
DM mass larger than $10^{11}M_{\sun}h^{-1}$.

Guo et al. (2011) calculated the two-point correlation
function and analysed galaxy clustering for MS
and for the Millennium II (Boylan-Kolchin et al. \citeyear{boylan2009})
simulations at small scales and found out that for stellar masses $M_*$
$9.77 < \log M_* < 10.77$ and for scales $<1$ $h^{-1}$Mpc, galaxies
are
more clustered in simulations than in observations (SDSS DR7).
Orphan galaxies account for almost 
half of all cluster members with $M_*>10^{10}M_{\sun}h^{-1}$ and 
thus galaxy abundance in clusters is underpredicted in MS
They note that the reason may be that galaxy disruption is not
  modelled properly in SAM or then the slightly wrong $\sigma_8$-value in MS
that follows DM halo and subhalo evolution may cause this difference.
In Bertone's data (used in this paper), about 18 percent of the M$\le -18$  galaxies are classified orphans (type 2). 

\subsection{Group rms velocities}
\begin{figure*}
 \centering
 \resizebox{\hsize}{!}{\includegraphics*[angle=0]{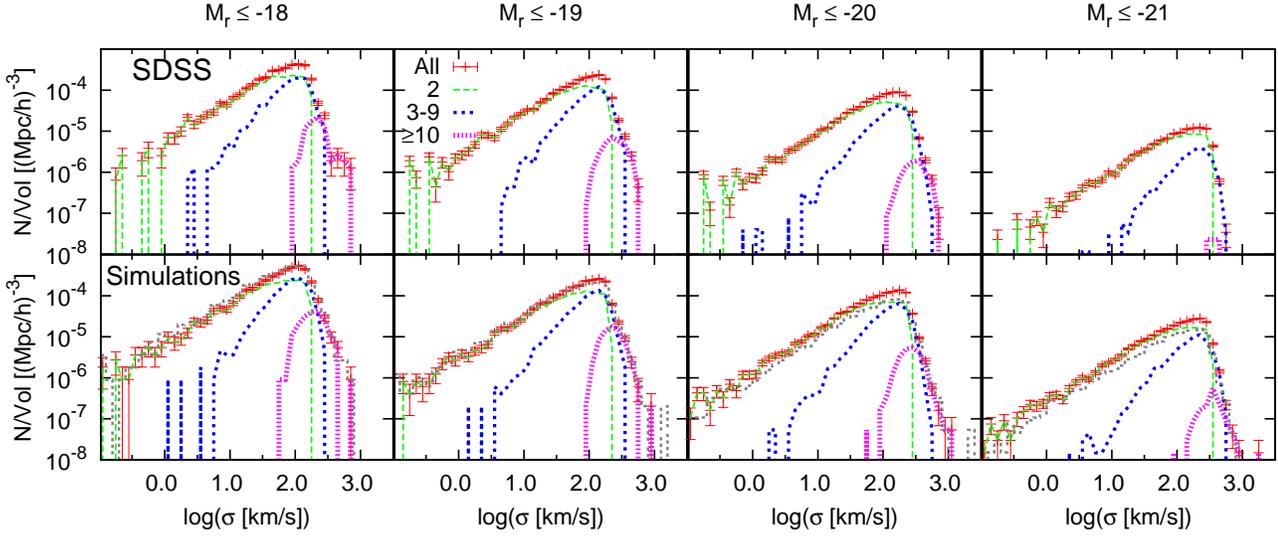}}
 \caption{Comparison of rms velocity distributions for SDSS
   and for Bertone SAM based on MS. The total distribution is shown
  as red points and 
the Poisson errors are shown in the bins. The modified distributions for galaxy pairs
(green dashed line), groups with $N_{\mathrm{gal}}=$3--9 members (blue dotted line) and
groups with $N_{\mathrm{{gal}}}\ge 10$ members (magenta dotted line) are also
given.}
\label{sigmav}
\end{figure*}

To compare dynamical properties of  groups, we calculated the
rms velocities
 for groups:               
\begin{equation} 
\sigma_v = \left(\frac{\sum_{i=1}^{N}|\vec{v}_i-\vec{v}_{\mathrm{mean}}|^{2}}{n-1}\right)^{1/2} ~, 
\end{equation} 
where $\vec{v}_i$ is the velocity of the member galaxy $i$, $n$ is the number of galaxies in the group and $\vec{v}_{\mathrm{mean}}$
is the mean velocity of galaxies.
We use this estimator for the group velocity differences because it is also used in the Tago et
  al. (2010) catalogue. For pairs and small groups $\sigma_v$ has no real physical meaning,
but it can still be calculated to have a complete sample for all groups. The distributions of $\sigma_v$ are shown in Fig. ~\ref{sigmav}.
There is an abundance of groups with the rms velocity close to
 100 km $\mathrm{s}^{-1}$ in all cases, and all the distributions agree very well
up to this point. For the $N_{\mathrm{gal}}=$3--9 and $N_{\mathrm{gal}}$$\ge 10$ groups the distributions
obtained for observations and simulations differ for
the two brightest samples. This is expected since the richness
distributions are also different for these cases, but the shapes of the $\sigma_v$ distributions are 
similar for all groups and samples.
If we compare the mean values of the $\sigma_v$ distributions (Table \ref{tab8}) we notice 
that the mean values for the mock groups and the SDSS
groups are almost the same within the error bars. For the $N_{\mathrm{gal}}=$3--9 groups
the mean value increases from $\sim 100$ km $\mathrm{s}^{-1}$ to $\sim
200$ km $\mathrm{s}^{-1}$ as we
move from the M $\le -18$ sample to the M $\le -21$ sample. For the $N_{\mathrm{gal}}$$\ge 10$
groups the mean value increases from $\sim 200$ km $\mathrm{s}^{-1}$ to $\sim 400$
km $\mathrm{s}^{-1}$, accordingly.
\begin{table}
 \caption{Mean values of $\sigma_v$ of
   galaxy groups for all $N_{\mathrm{gal}}=$3--9
   and $N_{\mathrm{gal}}$$\ge 10$ group distributions shown in Fig. ~\ref{sigmav} (in units of 
   km $\mathrm{s}^{-1}$).}
 \label{tab8}
 \centering
 \begin{tabular}{ccc}
 \hline \hline\\[-8pt]
Sample, $N_{\mathrm{{gal}}}$ & Simulations (Mean) & SDSS (Mean) \\[6pt]
 \hline
M $\le -18$, 3--9 & 97.5$\pm 1$  & 98.7$\pm 1$\\
$\ge10$	& 202$\pm 6$ & 229$\pm 9$\\
\hline
M $\le -19$, 3--9 & 113$\pm 1$ & 119$\pm 0.9$\\
$\ge10$	& 238$\pm 5$ & 258$\pm 7$\\
\hline
M $\le -20$, 3--9 & 144$\pm 1$ & 149$\pm 1$\\
$\ge10$	& 284$\pm 6$ & 320$\pm 8$\\
\hline
M $\le -21$, 3--9 & 183$\pm 2$ & 195$\pm 2$\\
$\ge10$	& 371$\pm 19$ & 360$\pm 30$\\
\hline
\end{tabular}
\end{table}

\subsection{Correlations between different galaxy group properties}

We have studied above the overall distributions of
different group properties. 
To understand the mutual dependencies of group properties we
have calculated the linear Pearson product-moment correlation coefficients $r$ 
between the main group properties: group luminosity,
$R_{\mathrm{vir}}$, $R_{\mathrm{max}}$ and $\sigma$. The correlations are calculated
separately for the observations and for the mock catalogues, as well as separately 
for the two group populations with $N_{\mathrm{gal}}=$3--9
members or with $N_{\mathrm{gal}}$$\ge 10$. Galaxy pairs are ignored, since some
 of the properties are not properly defined for them.
In all samples when the correlation coefficient is $r>0.15$, it is significantly different from
zero (with $p < 0.05$), since the degrees of freedom $(N-2)$ are always large for all
$r$-values referred to in the text (see Table \ref{tab1} for the SDSS data). 
To estimate the precision of sample statistics we use the so
  called jackknifing method. 
From the full data sets we derive 100
  subsets of available data and calculate correlation coefficients for each
  subsets. The subsets are selected so that, for the
  $N_{\mathrm{gal}}=$3--9 groups we start initially from the groups with $N=$4--10
  members and remove randomly one member from each group. Similarly
  for $N_{\mathrm{gal}}$$\ge 10$ groups we start from the groups with $N \ge 11$
  members. Then all
  group properties ($L_{\mathrm{g}}$, $R_{\mathrm{vir}}$, $R_{\mathrm{max}}$ and $\sigma$) are
  calculated again 100 times. This gives us the distribution of
  correlation coefficient values. In the following
  results we give the mean value and the standard deviation of the
  correlation coefficient distribution. 
This analysis shows that our results are not very sensitive for 
resampling, which indicates that groups are 
mainly real dynamical systems.

In general, calculations show that
correlations are usually very similar for different volume-limited
samples, but for those parameters that are clearly correlated, we give
a complete sample.
As expected, the group virial radius $R_{\mathrm{vir}}$ and the group size $R_{\mathrm{max}}$ are
always very well correlated (Table \ref{tab8}) with $r=$0.50--0.82
for Bertone SAM data and $r=$0.68--0.82 for SDSS data. The
correlations are systematically larger for the observed groups than for
the mock sample. There is no clear difference between $N_{\mathrm{gal}}=$3--9 groups
and $N_{\mathrm{gal}}$$\ge 10$ groups.
For groups in general, $R_{\mathrm{vir}}$ is the lower
limit of $R_{\mathrm{max}}$ and typically $R_{\mathrm{max}} \sim$1.5--2$R_{\mathrm{vir}}$ for the
$N_{\mathrm{gal}}=$3--9 groups and $R_{\mathrm{max}} \sim$2--4$R_{\mathrm{vir}}$ for the
$N_{\mathrm{gal}}$$\ge 10$ groups. 

\begin{table}
 \caption{Correlation coefficients for correlations between $R_{\mathrm{vir}}$ and  $R_{\mathrm{max}}$.
   Galaxy groups with $N_{\mathrm{gal}}=$3--9
   and $N_{\mathrm{gal}}$$\ge 10$ group members are included in the
   analysis. Errors given in the table for the mean values are
   $1\sigma^{-1}$ standard deviations of the mean value distributions obtained with the
   jackknifing test.}
 \label{tab8}
 \centering
 \begin{tabular}{ccc}
 \hline \hline\\[-8pt]
Sample, $N_{\mathrm{gal}}$ & Simulations & SDSS \\[6pt]
 \hline
M $\le -18$, 3--9 & 0.664 $\pm$ 0.013  & 0.749 $\pm$ 0.010\\
$\ge10$	& 0.664 $\pm$ 0.010 & 0.818 $\pm$ 0.012\\
\hline
M $\le -19$, 3--9 & 0.629 $\pm$ 0.013 & 0.729 $\pm$ 0.008\\
$\ge10$	& 0.644 $\pm$ 0.012 & 0.718$\pm$ 0.014\\
\hline
M $\le -20$, 3--9 & 0.657 $\pm$ 0.011 & 0.736$\pm$ 0.007\\
$\ge10$	& 0.823 $\pm$ 0.007 & 0.680$\pm$ 0.016\\
\hline
M $\le -21$, 3--9 & 0.630 $\pm$ 0.013 & 0.744$\pm$ 0.017\\
$\ge10$	& 0.504 $\pm$ 0.021 & - \\
\hline
\end{tabular}
\end{table}

\begin{table}
 \caption{Correlation coefficients for correlations between $L_{\mathrm{g}}$ and $R_{\mathrm{max}}$.
   Galaxy groups with $N_{\mathrm{gal}}=$3--9
   and $N_{\mathrm{gal}}$$\ge 10$ group members are included in the
   analysis. Errors given in the table for the mean values are
   $1\sigma$ standard deviations of the mean value distributions obtained with the
   jackknifing test.}
 \label{tab9}
 \centering
 \begin{tabular}{ccc}
 \hline \hline\\[-8pt]
Sample, $N_{gal}$ & Simulations & SDSS \\[6pt]
 \hline
M $\le -18$, 3--9 & 0.212 $\pm$ 0.016  & 0.348 $\pm$ 0.015\\
$\ge10$	& 0.878  $\pm$ 0.0038 & 0.756 $\pm$ 0.010\\
\hline
M $\le -19$, 3--9 & 0.234 $\pm$ 0.018 & 0.405 $\pm$ 0.011\\
$\ge10$	& 0.703 $\pm$ 0.0080 & 0.572 $\pm$ 0.013\\
\hline
M $\le -20$, 3--9 & 0.337 $\pm$ 0.010 & 0.443 $\pm$ 0.0085\\
$\ge10$	& 0.783 $\pm$ 0.0036 & 0.461 $\pm$ 0.014\\
\hline
M $\le -21$, 3--9 & 0.271 $\pm$ 0.015 & 0.505 $\pm$ 0.016\\
$\ge10$	& 0.805 $\pm$ 0.011 & - \\
\hline
\end{tabular}
\end{table}

\begin{table}
 \caption{Correlation coefficients for correlations between $L_g$ and $\sigma$.
   Galaxy groups with $N_{\mathrm{gal}}=$3--9
   and $N_{\mathrm{gal}}$$\ge 10$ group members are included in the
   analysis. Errors for the mean values are $1\sigma$ standard
   deviations and those are obtained with the
   jackknifing test.}
 \label{tab10}
 \centering
 \begin{tabular}{ccc}
 \hline \hline\\[-8pt]
Sample, $N_{\mathrm{gal}}$ & Simulations & SDSS \\[6pt]
 \hline
M $\le -18$, 3--9 & 0.186 $\pm$ 0.019  & 0.235 $\pm$ 0.016\\
$\ge10$	& 0.596 $\pm$ 0.0065 & 0.749 $\pm$ 0.0084\\
\hline
M $\le -19$, 3--9 & 0.196 $\pm$ 0.018 & 0.296 $\pm$ 0.010\\
$\ge10$	& 0.376 $\pm$ 0.012 & 0.486 $\pm$ 0.012\\
\hline
M $\le -20$, 3--9 & 0.190 $\pm$ 0.010 & 0.320 $\pm$ 0.0086\\
$\ge10$	& 0.658 $\pm$ 0.0050 & 0.473 $\pm$ 0.011\\
\hline
M $\le -21$, 3--9 & -0.0553 $\pm$ 0.019 & 0.229 $\pm$ 0.018\\
$\ge10$	& 0.968 $\pm$ 0.012 & - \\
\hline
\end{tabular}
\end{table}

On the other hand, there is no correlation between $R_{\mathrm{vir}}$ and
$\sigma$ for all groups. 
The correlation is larger, but still weak between $R_{\mathrm{max}}$ and $\sigma$: 
$r=$0.1--0.2 and $r=$0.1--0.7, for $N_{\mathrm{gal}}=$3--9 and $N_{\mathrm{gal}}$$\ge 10$,
respectively. The scatter between different samples is large, but
there is no systematical difference between the mock data and observations.
The dynamical state of the group and the group mass describe very diverse
galaxy environments and therefore the correlation between the group size
indicators and the velocity distribution is rather weak. 

Lastly, we studied the correlations between the group luminosity, that is related
to the group mass, and  $R_{\mathrm{max}}$ and $\sigma$ (Tables \ref{tab9} and \ref{tab10}).
For intermediate groups ($N_{\mathrm{gal}}=$3--9) the
correlation is weak:  $r=$0.21--0.50 between $L_g$-$R_{\mathrm{max}}$ and
$r=$0.0--0.32 between $L_g$-$\sigma$. For SDSS data the correlations
are systematically stronger.  
For $N_{\mathrm{gal}}$$\ge 10$ groups there is substantial correlation
$r=$0.46--0.88 for $L_g$-$R_{\mathrm{max}}$, but for the mock data the
correlation are larger than for SDSS data.
For correlations between $L_g$-$\sigma$ $r=$0.38--0.97, but the
variations are so large that there are no systematic differences.
We notice that weak correlation between luminosity
and rms velocity for $N_{\mathrm{gal}}=$3--9 groups may indicate that these groups are not usually
virialized. For rich groups larger correlation coefficients
supports higher degree of virialization. For example, by comparing
simulations and nearby groups of galaxies Niemi
et al. (2007) concluded that approximately 20 percent of nearby groups of galaxies are
not bound systems.
The strong correlation between $R_{\mathrm{vir}}$ and $R_{\mathrm{max}}$ makes their
distributions quite similar (Figs. 4 and 5). 
Since the correlations are stronger for rich
groups, these distributions are also more similar than for other
group populations.

\subsection{Radial distribution of galaxies in galaxy groups}

It is known that in galaxy clusters the observed galaxy
distribution follows very well the underlying DM distribution
(Carlberg et al. 1997, Biviano \& Girardi 2003, van
der Marel et al. 2000, Lin, Mohr \& Stanford 2004).
Many studies have shown that also satellite galaxies roughly follow
a NFW profile inside dark matter haloes, although in some studies
satellites are less centrally concentrated than the DM 
(e.g. Yang et al. 2005, Chen 2008, More et al. 2009).
However, in groups and in the systems that do not have a common DM
halo, the observed galaxies in groups
may have a different radial distribution.
We can analyse this by calculating 
radial number density distributions of galaxies in groups for the SDSS
data and for MS.

The observed radial velocities include position errors due
to the peculiar velocities of the galaxies in the group. These errors,
finger-of-gods, do not influence any other distribution than the
radial line-of-sight distributions of galaxies calculated from the group centre.
We correct this effect by following the procedure given in 
Liivam\"agi, Tempel and Saar (2012). For every galaxy in
the group there is the initial distance $d_{\mathrm{init}}$ that includes the redshift
distortion. We give the new distance $d_{\mathrm{new}}$ to the galaxy by using
$\sigma_v$ and the rms
projected distance $\sigma_d$. The mean distance
of the group members $d_{\mathrm{gr}}$ is also needed. Then the new distance is
calculated as:
\[
d_{\mathrm{new}}=d_{\mathrm{gr}}+(d_{\mathrm{init}}-d_{\mathrm{gr}})\frac{\sigma_d}{\sigma_v/H},
\]
where $H$ is the Hubble constant.
It should be noted that this correction is only statistical and
  $d_{\mathrm{new}}$ is not always the true 3D distance. However, for the
  comparison study this is an important correction that needs to be
  taken into account.
By using $d_{\mathrm{new}}$ and the sky coordinates we then calculate new position
vectors $\bmath{r_i}$ for the group galaxies.

It is not evident what is the ``true'' centre of the
galaxy group in the observational data. In our calculations the centre of the group refers to
the luminosity centre of the group member galaxies. This is calculated
in the same way as the centre-of-mass, but instead of masses we use
the luminosities of the galaxies. The centre of the group is then
\[
\bmath{R}=\frac{\sum_{i=1}^N L_i \bmath{r_i}}{\sum_{i=1}^N L_i},
\]
where $\bmath{R}$ is the position vector of the group centre, $L_i$ is
the luminosity of a galaxy and $\bmath{r_i}$ is its position vector.
The group-centric distance $\Delta=|\bmath{R}-\bmath{r_i}|$ between the galaxy and the group
centre can be then found. Usually the location of the
main galaxy (most
luminous) position is slightly shifted from the group luminosity centre. For
example, in Skibba et al. (2011) they noticed that the fraction of
massive groups in which the brightest galaxy is not the central galaxy
is very high $\sim 40\%$. Similar results have also been reported in Coziol
et al. (2009) and Einasto et al. (2012).

\begin{figure*}
 \centering
 \resizebox{\hsize}{!}{\includegraphics*[angle=0]{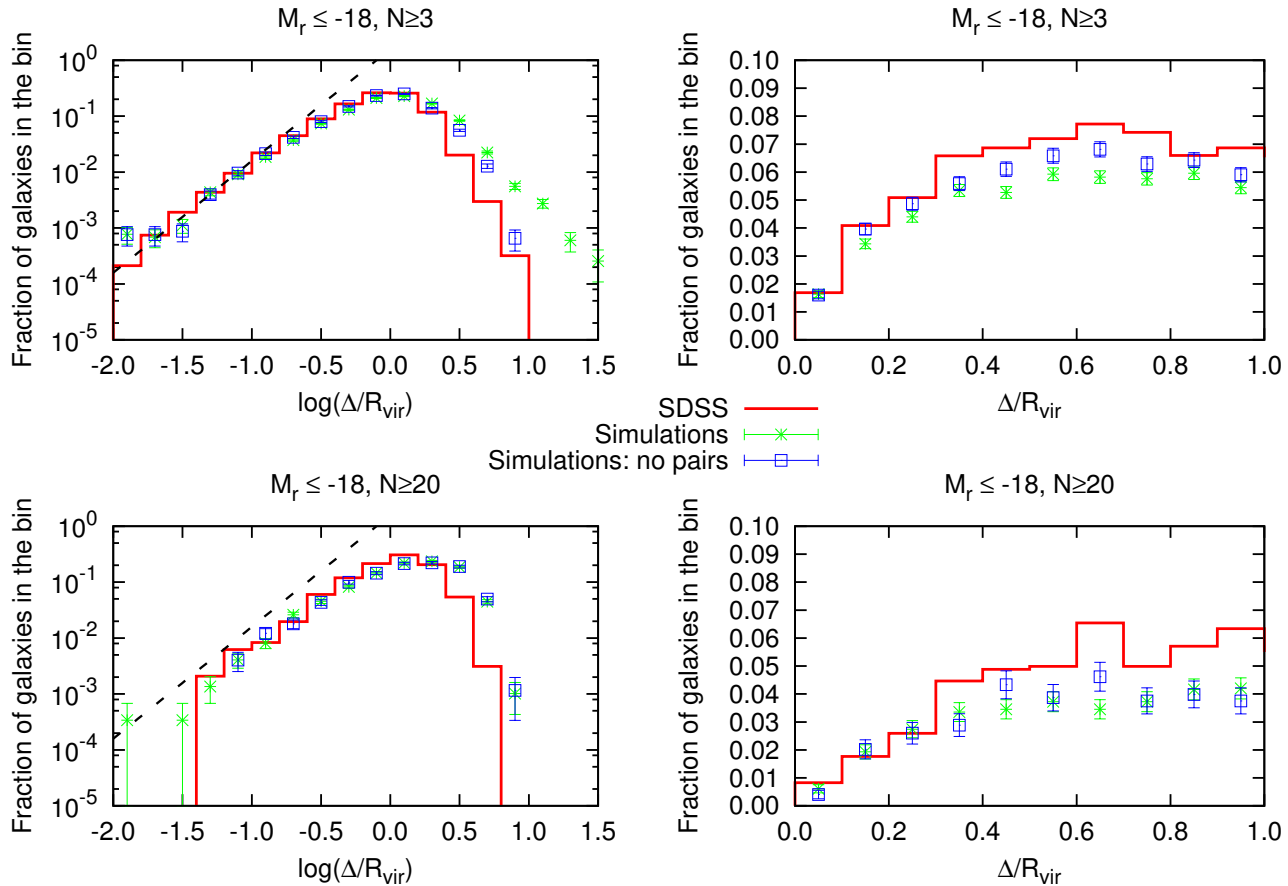}}
 \caption{Radial number distribution of galaxies as a function of the
   normalised distance $\Delta/R_{\mathrm{vir}}$ from the
   luminosity centre. All galaxy groups are averaged together and
   scaled with respect to their virial radii. Black dashed lines
   show the distributions for the uniform spatial density of
   galaxies. The left panels show the distributions up to $\Delta \sim 30
   R_{\mathrm{vir}}$ and the right panels show only the inner regions of the
   groups. The upper panels are for groups with $N \ge 3$ and the lower panels show
 only rich groups with $N \ge 20$. ``No pairs''-data points are for
 the modified mock data (see Sec.~3.4).}
\label{raddist}
\end{figure*}

To study the spatial distribution of galaxies in groups 
we calculated the 3D galaxy number density distributions for all groups with $N_{\mathrm{gal}} \ge 3$, showing
the number of galaxies at a certain normalised distance $\Delta/R_{\mathrm{vir}}$ from the group luminosity
centre (Fig. ~\ref{raddist}). 
The results are given only for the M $\le -18$ samples, because the
distributions are similar in other samples and this sample includes the
richest groups (see Fig. 3).
For comparison also the distributions
for very rich groups ($N_{\mathrm{gal}} \ge 20$) are given 
(only 5 in the SDSS sample). In addition to the normal mock
catalogue, we also show the distributions for the modified mock
catalogue from which all close pairs are removed in the same way as in Sec.~3.4.
Fig.~8 shows that after this correction (blue squares in the figure) the observed and theoretical
distributions are closer to each other.
Without the correction, the observed galaxy distribution
significantly differs from the galaxy distributions in the simulation.
The left panels show the radial distributions up to $\Delta \sim 30
R_{\mathrm{vir}}$ in the logarithmic scale and the right panels show only the inner
parts of groups (up to the virial radius) in the linear scale. In some groups that have
  very close galaxy pairs $R_{\mathrm{vir}}$ may be very small $\sim
  10^{-3}-10^{-4}h^{-1}$Mpc and therefore $\Delta/R_{\mathrm{vir}}$ can have
  large values. 
We note that the values this small are not realistic, since the
  typical galaxy size is in order of 10kpc. 
This shows that in some cases (especially for groups with only a few
members) the group $R_{\mathrm{vir}}$ is not the real virial radius.

The dashed lines in Fig.~8 show an uniform distribution of galaxies and all
groups follow this curve up to $\sim 1 R_{\mathrm{vir}}$. After this point
the observations and simulations start to notably deviate from the
uniform galaxy distribution. In the observations there are remarkably less galaxies outside the
virial radius at the outskirts of the groups. If close galaxy pairs are
removed, the agreement outside
the virial radius (as defined in this study) is much better, but
in simulations there are still more galaxies
in the outskirts of the groups than in the observations.
The group richness does not have a notable effect on the distributions,
since the upper panels and lower panels are qualitatively similar.
By using another observational group catalogue and a different mock group catalogue, Snaith et
al. (2011) noticed that SAMs produce radial distributions of galaxies
which is more centrally concentrated than in the
observations. Although our results and theirs are not directly
comparable, we observe a similar small effect (upper left panel). 
The difference would be more evident if the distributions are normalised.

Lastly, we studied the spatial location of the most luminous galaxy
in the group. For $N_{\mathrm{gal}}$$\ge 10$ groups, 
in observations the mean distance between the luminosity
centre and the brightest galaxy is $\sim 0.15 \pm 0.01$ $h^{-1}$Mpc for the M
$\le -18$ sample and $\sim 0.52 \pm 0.10$ $h^{-1}$Mpc for the M
$\le -21$ sample. In simulations the mean distance is the same for the
M $\le -18$ sample,  $\sim 0.13 \pm 0.01$ $h^{-1}$Mpc, but notably smaller for
the M $\le -21$ sample, $\sim 0.15 \pm 0.01$ $h^{-1}$Mpc. Also, mean
luminosities of the brightest galaxies are systematically 2--5 times larger in
the simulations compared with observations. Since the samples used in
this analysis can be connected to the mean mass of the groups (see
Table 4) we conclude that SAM
used in this analysis produces too bright galaxies close to the centre
of the groups in the massive galaxy groups. 

In general, our galaxy distributions cannot be
  directly compared with other galaxy density distribution studies
  (e.g. Yang et al. 2005, Chen 2008, More et al. 2009).
Our normalized galaxy density distribution is based on the
statistically corrected 3-D positions and this makes the distribution
more uniform than what it really is.
Our distribution does not directly correspond the true galaxy density
profile. Still, we can analyse the differences between the observed
and SAM data that are processed in the same way.

One explanation for the observed difference may be due to physical processes influencing
galaxy evolution that are not modelled correctly by SAMs.
There are many processes that may influence the results at the
outskirts of groups and cause the overabundance of
galaxies. Although many physical processes are effective at the galaxy
cluster scale, it has been shown that the same processes can be
effective also in low-velocity dispersion systems 
(e.g., Zabludoff \& Mulchaey 1998, Weinmann et al. 2006a).
The ram pressure stripping (Gunn \& Gott 1972), a cut-off of gas accretion onto galaxy
discs by different processes (strangulation, Larson et al. 1980,
Balogh, Navarro and Morris 2000, Croton et al. 2006) and
high-velocity close encounters of satellite galaxies called harassment
(Moore et al. 1996) are all examples of
processes that can change the $r$-band luminosities of galaxies
(visibility in the mock data) and therefore
affect the shape of the distribution. 

Differences between Durham and Munich SAMs
are studied in detail by Contreras et al. (2013).
They show that the Durham and Munich models produce  
a different spatial distribution 
of satellite galaxies within the main halo. 
The Durham satellite galaxies are more clustered at 
small scales while the spatial distribution of the resolved subhaloes
is more extended in the Munich model.  
Moreover, the Munich model produces about 
three times more satellites with 
resolved subhaloes than the Durham one. 
 Contreras et al. (2013) concluded that the reason for this is 
the difference between subhalo/satellite merger procedures in the Munich and Durham models.

\section{Conclusion and Discussion}
Although galaxy groups are the most common environment for galaxies, this 
very diverse population from loose associations of a few galaxies up
to systems with several tens of galaxies is relatively poorly studied. 
An important result obtained in many studies is that physical properties of galaxies
(colours etc.) are different in different types of groups 
(e.g. Yang et al. 2008, Weinmann et al. 2006a, Weinmann et al. 2006b, 
Mendel et al. \citeyear{mendel2011}).
Although DM drives the formation of the large-scale structure, 
the models of formation of galaxies are far from completely understood.
This problem has been stressed by several authors and disagreements
between the mock catalogues and observed catalogues have been reported.
For example, \cite{robotham} found a remarkable
deficit in the number of observed high multiplicity groups compared to the mocks. 
They also found that there were significantly less compact
groups in the observed data. 

However, it should be remembered that
the mock catalogues and the results derived from them are cosmology
dependent. In Yang et al. (2005) and van den Bosch (2005) it was shown
that a cosmology with $\sigma_8 \sim 0.7$ gives better agreement
between simulations and observations for high multiplicity groups than
$\sigma_8 \sim 0.9$ used in MS.
Using the rescaling technique of Angulo \& White (2010), Guo et al. (2013) 
scaled MS based on WMAP1 cosmology (Bennett et al. 2003) to 
WMAP7 cosmology (Komatsu et al. 2011). 
They concluded that there is no considerable effect of different cosmologies   
for halo mass functions at least up to $z=3$. The 
amplitude parameter $\sigma_8$ is lower in the WMAP7 cosmology, but the matter density 
$\Omega_m$ is higher for WMAP7. The net result is that 
these two effects cancel each other. 
They summarize, that different cosmologies (WMAP1 or WMAP7) 
produce only small differences in galaxy properties. 
In Guo et al. (2013) fig.~6, they show that their SAM (Munich model),   
updated for WMAP7 cosmology, slightly increases the abundance of luminous galaxies.

Observationally and physically the definition of a galaxy group is not trivial, since they represent a
wide mass range of systems from individual galaxies and their satellites to large clusters of galaxies.
There are many methods and algorithms to identify galaxy groups and
clusters (e.g. Hugra \& Geller 1982, Yang et al. 2005b, Tago et al. 2010) that have been used to extract
different group catalogues. This makes it very complicated to
compare the group properties in different catalogues in the same way
and the group catalogues need to be studied separately.
We have to also remember that observational galaxy catalogues are never ideal. Different observational
constraints need to be taken into account and direct comparison may
contain observational limitations.
We have tried to eliminate some of these biases by using 
volume-limited galaxy samples. This allowed us to compare the number densities of the 
groups and to compare group properties directly with simulations. 

In our study we compared the mock group catalogues with the SDSS group catalogues, both 
constructed by using the same algorithm.
For the most part these two group catalogues agree well.
First, our comparison showed that the richness distributions agree well between the 
simulations and SDSS. The numbers of pairs and loose groups match very
well, but simulations give richer groups. 
Also the $\sigma_v$ distributions are very similar, confirming the 
idea that the velocity distribution is not sensitive to the group
definition. The rms velocity $\sigma_v$ does not reflect efficiently the differences in
the group properties. 
Clear discrepancies are seen in the group 
size and virial radius distributions. 
Also, there are clear differences in the spatial galaxy distributions. 
In simulations there are more galaxies 
beyond the virial radius than in the observations, if no modifications
are done to the mock catalogue. 
The agreement is much better if we
remove all close galaxy pairs from the mock catalogue, mimicking the observational bias.
Also, different physical processes that are not modelled properly by SAM
may play a role in this discrepancy.
Another reason for this discrepancy might be the fact that 
the connection between the DM haloes and galaxies is not trivial.
Some subhaloes may not have an observed counterpart as suggested 
by \cite{gao}. \cite{guo} also found that galaxy
distributions in rich clusters agree between their simulations and
observations only if galaxies without DM subhaloes (orphan galaxies) are
included. This can be problematic in SAMs, where DM halo
mergers and masses are followed during the simulations and galaxy
modelling is based on the DM halo evolution.

Finally, to check how the overabundance of bright galaxies in the simulated
catalogues affects our results, we removed all bright galaxies with
$M_r-5\log h \le -22$ from the
simulated data  (see Fig. 1),
produced all the mock group property distributions again and studied how the
distributions differ from the original ones. The distributions were
surprisingly similar with the original distributions 
and all the values in the histograms were within
the error bars, except for the $M_r \le -21$ sample. Here marginal differences
were seen and the agreement with the observations was
slightly better, but all the conclusions and general results given in
the paper are still valid, even if the brightest galaxies are totally
excluded from the simulations.

The main results of this study are as follows:
\begin{enumerate}
\item We have calculated the statistical properties of the SDSS-DR7 group
  catalogues. The results are listed in Tables 1, 2, 5--8.  
\item In the Bertone SAM data the luminosity function of galaxies
  differs notably from observations for $M_r-5\log h \le -21.5$, being biased
  towards higher luminosities. 
\item 
For SDSS data the richness distributions follow simple
power laws and the calculated logarithmic least-square slope values are
$-2.02\pm0.18$, $-2.12\pm0.17$, $-2.26\pm0.19$ and $-3.29\pm0.24$ for
$M_r \le -18$, $-19$, $-20$ and $-21$, respectively. SAM results from the
Bertone and Font data give significantly different results from
observations. There are too many rich and luminous groups in SAM data
compared with observations.
\item The SDSS group catalogue includes incompleteness at small galaxy
  separations. This is especially
  important for galaxy
  pairs of small sizes. 
For $\log R_{\mathrm{max}}[h^{-1}\mathrm{Mpc}]$ these limitations start from $-1.46$,
$-1.28$, $-1.11$ and $-0.0955$, for the volume-limited samples $M_r \le -18$,
$M_r \le -19$, $M_r \le -20$, $M_r \le -21$, respectively, and grow for smaller sizes.
Since $R_{\mathrm{vir}}$ and $R_{\mathrm{max}}$ are strongly correlated, the same limitation
is a problem also for the $R_{\mathrm{vir}}$ distribution. 
\item For all samples the correlation between $R_{\mathrm{vir}}$ and the rms velocity $\sigma$ is
either weak or non-existent. Excluding galaxy pairs, in observations the group
luminosity is weakly correlated with $R_{\mathrm{vir}}$,  $R_{\mathrm{max}}$ and $\sigma$.
\item In the Tago et al. (2010) catalogue there are 
  less galaxies 
at the outskirts of a group than in
simulations. In the inner regions the galaxies have an almost uniform number
density for all groups. The agreement is better if the
observational galaxy pair bias is taken into account. However, this
distribution does not correspond to the true galaxy density
profile, since in our comparison we use a statistical 3-D radius that
makes the distribution more uniform than it really is.
\end{enumerate}

Summarizing, the comparison between the mock group
catalogue and observations reveal interesting differences. 
Some of these differences are due to the problems in the properties of galaxies
generated by semi-analytical methods (SAMs). Some differences
can be associated with the observational limitations in the SDSS
galaxy data. In future studies, we will analyse how tight is the
connection between the underlying DM halo and the galaxy group. Do all
observed groups have a main DM halo or are they in some cases just loose
connections of separate DM haloes?

\section*{Acknowledgements}
This study was supported by the Finnish
Academy funding, the Turku University Foundation, the Estonian Science Foundation
grants No. 8005, 7765, 9428 and MJD272, the Estonian Ministry for Education
and Science research project SF0060067s08, and by the European Structural
Funds grant for the Centre of Excellence "Dark Matter in (Astro)particle
Physics and Cosmology" TK120. This work has also been supported by
ICRAnet through a professorship for Jaan Einasto. We thank Chris Flynn
and Sarah Bird for all valuable corrections and comments about the paper.

\label{lastpage}
\end{document}